\documentclass{aastex631}

\shorttitle{Spectral Identification for White Dwarf from Gaia EDR3}
\shortauthors{Xiao Kong et al.}
\graphicspath{{./}{figures/}}

\begin{document}

\title{Identification of White Dwarf from Gaia EDR3 via Spectra from LAMOST DR7}

\correspondingauthor{A-Li Luo}
\email{lal@nao.cas.cn}

\author[0000-0001-8011-8401]{Xiao Kong}
\affil{Key Laboratory of Optical Astronomy, National Astronomical Observatories, Chinese Academy of Sciences, \\
	Beijing 100012, China}

\author[0000-0001-7865-2648]{A-Li Luo}
\affiliation{Key Laboratory of Optical Astronomy, National Astronomical Observatories, Chinese Academy of Sciences, \\
	Beijing 100012, China}
\affiliation{School of Astronomy and Space Science, University of Chinese Academy of Sciences, Beijing 100049, China}



\begin{abstract}

We cross-matched 1.3 million white dwarf (WD) candidates from Gaia EDR3 with spectral data from LAMOST DR7 within $3^{\prime\prime}$. Applying machine learning described in our previous work, we spectroscopically identified 6 190 WD objects after visual inspection, among which 1 496 targets were firstly confirmed. 32 detailed classes were adopted for them, including but not limited to DAB and DB+M. We estimated the atmospheric parameters for the DA and DB type WD using Levenberg- Marquardt least-squares algorithm (LM). Finally, a catalog of WD spectra from LAMOST was provided online.

\end{abstract}

\keywords{white dwarfs -- methods: data analysis -- techniques: spectroscopic -- catalogs}


\section{Introduction} \label{sec:intro}

WDs are assigned to several subtypes according to the major components of the surface atmosphere.
If normally broad Balmer lines appear in a spectrum, it is DA WD.
Similarly, He {\small \bf I}, He {\small \bf II}, Ca {\small \bf II} H \& K and C line for DB, DO, DZ, and DQ respectively \citep{2013ApJS..204....5K}. 

Approximately 1.3 million targets were selected from {\it Gaia} Early Data Release (EDR) 3 \citep{2021A&A...649A...1G}, considered to be the candidates of WD \citep{2021arXiv210607669G}.
These candidates were derived by several selection criteria in absolute magnitude, color, etc. from {\it Gaia} EDR3 catalog.
The samples of spectroscopically confirmed SDSS WDs were also adopted to calculate probabilities of being a WD ($P_{\rm WD}$).

Up to nearly 7 million objects, LAMOST \citep{2015RAA....15.1095L} released its 7th data product (Data Release 7, DR7) that included more than 10 million low resolution spectral data.
In this research, we aimed to identify WD candidates of {\it Gaia} EDR 3 using spectra from LAMOST DR7.

\section{Data Selection} \label{sec:data}

We performed a cross-match of WD candidates with LAMOST spectral data onto equinox 2000 and epoch 2000 within $3^{\prime\prime}$ utilizing formula \ref{eq:dis}, where $\alpha$ and $\delta$ represent right ascension and declination respectively.

\begin{equation}	\label{eq:dis}
d = \arccos[\cos \delta 1 \cos \delta 2 \cos(\alpha 1-\alpha 2)+\sin \delta 1 \sin \delta 2]
\end{equation}

Considering a LAMOST fiber diameter of $3^{\prime\prime}$ \citep{2015RAA....15.1095L}, the cross radius $d$ was also restricted to $3^{\prime\prime}$.
The accuracy of fiber positioning, on the other hand, had been determined to be no more than $1.5^{\prime\prime}$ on average \citep{2014SPIE..9149....1}.
We start the cross-match procedure based on sky coordinates by using both radii.
The conclusions of this research were based on radius $3^{\prime\prime}$, while a specific field presented in our catalog was used to mark the data if it was more than $1.5^{\prime\prime}$ away from any source of {\it Gaia}.

Moreover, these data that had positive values of offset were dismissed as the inconsistent between observation position and the coordinates from the input catalog.
We then derived 18 232 spectra.
Among all the 1.4 million WD candidates from {\it Gaia}, there existed 12 046 that corresponded with LAMOST spectra.

We also inspected 5.8 million spectra of SDSS DR16 \citep{2020ApJS..249....3A} and calculated the distances between them and our data sample.
Half of the objects of LAMOST's sample were not observed by SDSS yet, they needed to be verified even more.

\section{Identification} \label{sec:identity}

We adopted the same procedure described in \cite{2018PASP..130h4203K} to remove those that had little evidence of spectral lines.
9 496 spectra remained after the machine learning process.
Afterwards, we inspected all the results and rejected those that had little evidence to be a WD and maintained 8 465 spectra of 6 190 objects.
The number of targets would reduce to 6 045 if the radius was fixed at $1.5^{\prime\prime}$.
Meanwhile, these small samples of spectra that located between $1.5^{\prime\prime}$ and $3^{\prime\prime}$ exhibited no characteristics distinct from those within the radius of $1.5^{\prime\prime}$.

During the inspection, all the WD spectra were classified into detailed types, including but not limit of DAB and DBZ.

Considering the $P_{\rm WD}$ from \cite{2021arXiv210607669G}, about 75\% of the spectral confirmed WD objects had values greater than 0.9.
As for the other spectra that may not be WDs, the $P_{\rm WD}$ was relatively lower.
Around 68\% of them were less than 0.5, while 23\% larger than 0.9.
This is partly due to poor quality that we were unable to discover any line features in a spectrum and discard it from our sample.
In the other hand, they might be DC WD that could not be verified by spectra alone.

52 objects with 59 spectra were composed of 2 stellar systems, usually a WD plus a late-type star.
We utilized LAMOST 1D Pipeline to recognize the other components by template fitting and found the majority of them WD plus M.

Interestingly, 64 cataclysmic variables (CVs) \citep{1995cvs..book.....W}, with 49 had obvious double peak structure among their emission Balmer or He lines, were intermixed in our sample.
Most of them had traces of helium in their spectra.

\section{Stellar Parameters} \label{sec:param}

Applying spectral templates for DA and DB \citep{2010MmSAI..81..921K}, we calculated atmosphere parameters of the WD spectra.

In the beginning, all spectral samples were moved to rest-frame relying on the redshift derived from LAMOST 1D Pipeline.
Following the determination of best-fit model, LM  were adopted to estimate $T_{\rm eff}$, $\log g$ ~and their uncertainty.
The depth of lines of several spectra with relatively low signal-to-noise ratio, though, were almost the same intensity as noise.
The parameters of these data went out of bounds of the template scope and were to reset to -9999 manually.

\section{Summary}

We present a catalog of spectral confirmed WDs from LAMOST DR7 based on the candidates from {\it Gaia} EDR3.
The full {\it Gaia}-LAMOST spectroscopic sample catalog contains the main information (see Table \ref{tab:catalog}) of WDs and can be downloaded following link: \url{http://paperdata.china-vo.org/XiaoKong/LAMOST_DR7_WD.fits}

\begin{deluxetable*}{ll}
	\tablenum{1}
	\tablecaption{Format of the LAMOST DR7 Catalog of spectral confirmed WDs.\label{tab:catalog}}
	\tablewidth{0pt}
	\tablehead{
		\colhead{Heading} & \colhead{Description}
	}
	\startdata
	O{\scriptsize BSID}	&	Unique ID of a spectrum	\\
	S{\scriptsize OURCE\_ID}	&	Unique ID for this object in {\it Gaia} EDR3	\\
	W{\scriptsize D\_NAME}	&	LAMOSTJ+J2000 ra (hh mm ss:ss)+dec(dd mm ss.s), equinox and epoch 2000	\\
	RA		&	Right ascension [deg] of object	\\
	DEC		&	Declination [deg] of object		\\
	T{\scriptsize YPE}	&	Detailed classes for WDs from LAMOST	\\
	SNR		&	Signal-to-noise ratio of $u$, $g$, $r$, $i$ and $z$ filter	\\
	RV		&	Radial velocity [km/s] of object from template fitting		\\
	$T_{\rm eff}$	&	Effective temperature [K] from fitting the parameter model \\
	$\log g$	&	Surface gravity from fitting the parameter model	\\
	N\_{\scriptsize BIB}	&	Number of references that confirmed type (identify for the first time if 0)	\\
	R{\scriptsize AD}	&	1 means distance between this spectrum and {\it Gaia} source smaller than $1.5^{\prime\prime}$	\\
	\enddata
	\tablecomments{This catalog involves WD and CV information both. -9999 means the value can not be provided.}
\end{deluxetable*}

Starting from cross-matching spectral data of LAMOST with sources of {\it Gaia}, we identified 8 465 WD spectra, involving DA, DB, DO, DZ, etc.
Some other types, binary or CV for instance, were also noted.
These objects corresponded with 6 190 {\it Gaia} targets.
1 496 stars of our samples were spectral confirmed for the first time.
We then found the best model that fits a spectrum and estimated the atmosphere parameter using LM.

\begin{acknowledgments}
We thank J.K. Zhao and D. Koester for providing the parameter templates for DA and DB WD.
These models were made by D. Koester ranging from 5 000 K to 80 000 K and 7.0 to 9.5 for $T_{\text{eff}}$ and $\log g$ respectively.

This work has made use of data from the European Space Agency (ESA) mission
{\it Gaia} (\url{https://www.cosmos.esa.int/gaia}), processed by the {\it Gaia}
Data Processing and Analysis Consortium (DPAC,
\url{https://www.cosmos.esa.int/web/gaia/dpac/consortium}). Funding for the DPAC
has been provided by national institutions, in particular the institutions
participating in the {\it Gaia} Multilateral Agreement.

Guoshoujing Telescope (the Large Sky Area Multi-Object Fiber Spectroscopic Telescope LAMOST) is a National Major Scientific Project built by the Chinese Academy of Sciences. Funding for the project has been provided by the National Development and Reform Commission. LAMOST is operated and managed by the National Astronomical Observatories, Chinese Academy of Sciences.
\end{acknowledgments}

%

\vspace{5mm}
\facilities{{\it Gaia}, LAMOST}


\bibliography{ref}{}
\bibliographystyle{aasjournal}



\end{document}